# Experiments Indicate Regolith is Looser in the Lunar Polar Regions than at the Lunar Landing Sites

P.T. Metzger[1], S. Anderson[2], and A. Colaprete[3]

[1]Florida Space Institute, University of Central Florida, 12354 Research Parkway, Partnership 1 Building, Suite 214, Orlando, FL 32826-0650, USA; PH (407) 823-5540; email: philip.metzger@ucf.edu
[2]Department of Physics, University of Central Florida, 4000 Central Florida Blvd, Orlando, FL 32816, USA.
[3]Division of Space Sciences and Astrobiology, NASA Ames Research Center, Moffett Field, CA 94035, USA.

**ABSTRACT**
Since the Apollo program or earlier it has been widely believed that the lunar regolith was compacted through vibrations including nearby impact events, thermal stress release in the regolith, deep moon quakes, and shallow moon quakes. Experiments have shown that vibrations both compact and re-loosen regolith as a function of depth in the lunar soil column and amplitude of the vibrational acceleration. Experiments have also identified another process that is extremely effective at compacting regolith: the expansion and contraction of individual regolith grains due to thermal cycling in the upper part of the regolith where the diurnal thermal wave exists. Remote sensing data sets from the Moon suggest that the soil is less compacted in regions where there is less thermal cycling, including infrared emissions measured by the Diviner radiometer on the Lunar Reconnaissance Orbiter (LRO). Here, we performed additional experiments in thermal cycling simulated lunar regolith and confirm that it is an effective compaction mechanism and may explain the remote sensing data. This creates a consistent picture that the soil really is looser in the upper layers in polar regions, which may be a challenge for rovers that must drive in the looser soil.

## INTRODUCTION

The Apollo missions discovered that the lunar soil is more compacted than its own self-weight is capable of achieving, suggesting some geological process has actively compacted the soil:

> … the stress produced by self-weight is not sufficient to squeeze the soil to such high relative densities at shallow depth. At some time after deposition, the in situ lunar soil has either been exposed to higher stresses, or compacted by vibration, or a combination of the two processes…

> In fact, the actual in situ soil (at a depth of 50 cm) would had to have been compressed to a pressure greater than 100 kPa, or more than 100 times its present stress, to account for the measured density. On the other hand, if the soil had been vibrated after being deposited (i.e., by meteoroid impacts), the relative density might be 90% or more, and the density profile would be virtually uniform...
>
> To explain the extreme changes in relative density that occur in the top 30 cm of lunar soil, there must be a mechanism that both stirs up the surface and densifies the immediately underlying soil. Both of these effects are presumably caused by repeated meteoroid impacts. However, we do not know yet whether such impacts have a greater tendency to vibrate the soil or to stress it; the observed increases in relative density with depth could reflect either or both processes. (Carrier, Olhoeft and Mendel, 1991, pp. 503-504)

However, there are no published studies showing that meteorite impacts or other vibration sources on the Moon are adequate to perform this compaction. We should consider that impacts might produce as much "fluffing" of lunar soil as densification. Gault et al. (1974) performed hypervelocity impact experiments and published diagrams of the resulting subsurface deformation. Roughly scaling the volume of ejecta versus the volume of compressed soil, it can be seen the impacts created a significant net loosening, not a net compaction.

Shallow moonquakes are another possible candidate to produce adequate vibrations to compact lunar soil. They can be strong but they seem to be associated with the rims of ancient impact basins and are not homogeneously distributed around the Moon (Nakagawa, 2016) so the resulting vibrations at the surface may be inadequate at most locations. Gamsky and Metzger (2010) performed experiments compacting lunar simulant by vibration and showed that to achieve 1.95 g/cm$^3$ (approximately the asymptotic value measured beneath the Moon's surface), extremely high levels of vibration are required, as high as 0.23 g. (This is comparable to the acceleration in destructive earthquakes on Earth.) Considering the magnitude of events needed to cause adequate acceleration to compact the lunar soil, it is not *a priori* obvious that sufficiently strong and sufficiently many events have occurred with sufficient proximity to every patch of soil to produce the observed compaction, although we cannot rule it out.

On the other hand, there is evidence that the soil is less compacted toward the Moon's poles than at equatorial and mid-latitude locations. Hayne (2015) reported that global measurements by the Diviner Radiometer compared to thermal modeling indicate the lunar soil is systematically looser where there is less insolation, so it tends to be less compacted toward the poles. Hayne et al. (2017) reported soil in the range about ±60° latitude is remarkably homogeneous, but global maps over latitudes about ±70° (their Figs. 7 and 8) suggest soil becomes systematically looser just beyond ±60°. The impact of the Lunar Crater Observation and Sensing Satellite (LCROSS) in Cabeus crater, a Permanently Shadowed Region (PSR) near the south pole, resulted in a 300 ms delay prior to the observed infrared impact flash, and there was a suppression of the visible-band flash (Schultz et al., 2010). These were attributed to a highly porous (>70%) target material, providing ground truth of looser soil in a high latitude location.

Chen et al. (2006) performed thermal cycling experiments on an idealized granular material and showed thermal cycling is an effective compaction mechanism, an alternative to the vibrational mechanisms. Clegg and Metzger (Clegg, 2009) and Gamsky and Metzger (2010) performed similar experiments with lunar soil simulant and found thermal cycling effective to compact lunar soil within just several years of lunar rotations. This could help explain latitude dependence in the extremes of Diviner observations. However, it needs to be shown that unintended experimental vibrations were not causing the compaction during those experiments and that there is a correlation between the thermal wave amplitude and the resulting degree of compaction. In this paper we report new experiments to help address these. If thermal cycling without vibration is effective to compact soil, and if higher insolation (higher thermal wave amplitude) systematically makes the soil more compact, then it supports interpretation of the Diviner data that soil is marginally looser starting beyond ±60º latitude and LCROSS data that soil is extremely loose in a PSR. This is particularly important now since a rover is being designed for the Resource Prospector mission to visit a lunar polar region including driving through a PSR, and we must anticipate reasonably the soil conditions.

**THERMAL CYCLING EXPERIMENTS**

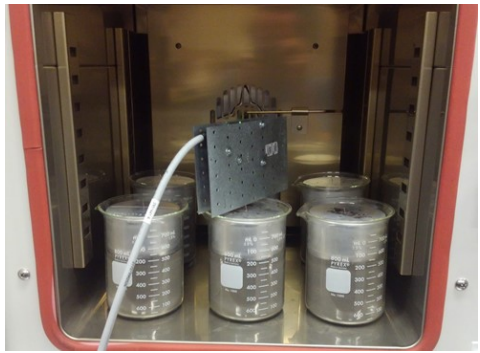

**Figure 1.** Beakers of JSC-1A Lunar Soil Simulant with weighted cups as overburden in the test oven. The front center cup has a laser displacement sensor placed on top to measure how far the overburden cup has settled.

This manuscript reports new thermal cycling experiments using JSC-1A lunar regolith simulant in a Jeio-Tech OF-02G programmable convection oven. This was not in vacuum but for these temperature profiles and the soil's permeability the pore pressure gradients should be sufficiently low to avoid disruption. For each temperature profile there were six samples consisting of 800 g of soil in an 800 mL

Pyrex beaker. Before beginning a new profile, the soil was de-compacted by stirring until it naturally rested at approximately 500 mL in the beaker for average bulk density $\rho$=1.6 g/cm$^3$. A flat implement acted as rake to flatten the simulant's top surface without further compaction, and a lidded aluminum cup (diameter=91mm, height=54 mm) was placed atop the simulant to provide a flat, highly reflective surface for the displacement measurements. The cup masses were 68.4 g ± 4%, which provides the same weight as a 0.66 cm layer of simulant in Earth's gravity. Four marks on the lid's surface indicated the points of measurement since the centroid was occupied by a lifting hook. Averaging over the four marked points would recover the height of the centroid.

As thermal cycling occurred the soil experienced a net deflation, i.e., its upper surface lowered as the soil became more compact. A Micro-Epsilon ILD 1302-50 laser displacement sensor measured the soil's deflation. It was mounted between two parallel plates to enable consistent placement on the sample beakers. To probe the effect of the different thermal wave amplitudes on the Moon, the samples were thermal cycled with four different amplitudes, $\Delta T$=187, 97, 67, and 37 ºC, which was accomplished at constant oven heating rate by using four different periods of the heating/cooling cycle, $P$= 6, 3, 2, and 1 hr, respectively. The sample temperatures were measured with a thermocouple embedded in each sample's center plus an infrared thermometer aimed at the surfaces of the samples, just before the heating phase and just before the cooling phase when the samples were at their coldest and hottest, respectively. The presence of the thermocouple disturbed the sample, and lifting an aluminum cup from the sample repeatedly to see the simulant's upper surface in infrared vibrated the sample unacceptably, so these temperature calibrations were performed only as a preliminary step. Samples were then re-prepared into the initial state but without the thermocouple, and subsequent thermal cycling was performed using the pre-calibrated oven period, measuring only the displacement of the top surface of the cup (without removing the cup) at particular intervals. This process was repeated for each of the four thermal amplitudes.

To isolate the vibration-induced compaction from the thermally-induced compaction, a control group was prepared consisting of four samples that were subjected to identical handling and measurements but with the oven deactivated. We initially found that despite careful handling the vibrational effects were unacceptably large. Shocks were caused by repeatedly opening and closing the oven door, taking samples in and out of the oven using handling devices that had to be attached and removed from the hot beakers, placing the samples onto a surface to take measurements, and setting the displacement sensor repeatedly onto each beaker's rim. Additional environmental shocks may have been transmitted through the facility structure into the oven. The authors redesigned the experimental methods twice, fabricated new handling hardware, procured a larger oven to enable taking measurements inside the oven to obviate removal of samples, and continuously repeated the experiments over the course of a year before obtaining statistically significant results. More expensive equipment is needed to improve further, including vibrational isolation from the facility, high temperature displacement sensors for use inside the oven with the door remaining closed, and vibration sensors to quantify the environment and correlate to particular compaction events.

Two methods of data reduction were employed. A few sample cups showed large, sudden jumps in the displacement measurement at a particular point in time. In the first method of data reduction, these sample cups were all removed entirely from the analysis. For most temperature amplitudes no more than 2 cups were eliminated, but the $\Delta T$=67 °C case had too many disturbances so the entire case was eliminated from the first data reduction. Also, biases often occurred beginning with the initial measurement, presumably because of the phasing of the thermal cycle and the expansion/contraction of the glass beaker relative to that of the soil. A constant value was added or subtracted to each curve to remove these initial biases. Ultimately, these constants had negligible effect on the fitting functions or the resulting analysis but they improved the plots' appearances. Figure 2 (left) shows the first data reduction's resulting curves of deflation for four cups at $\Delta T$=97 °C after the initial biases were removed. Figure 2 (right) shows fitted curves for three of the temperature ranges (only three for clarity). The control group ($\Delta T$=0 °C) deflated an average of ~0.2 mm over 120 cycles while the highest temperature case ($\Delta T$=187 °C) deflated an average of ~1 mm over 120 cycles. These measurements are 8.9 standard deviations apart demonstrating clear statistical significance.

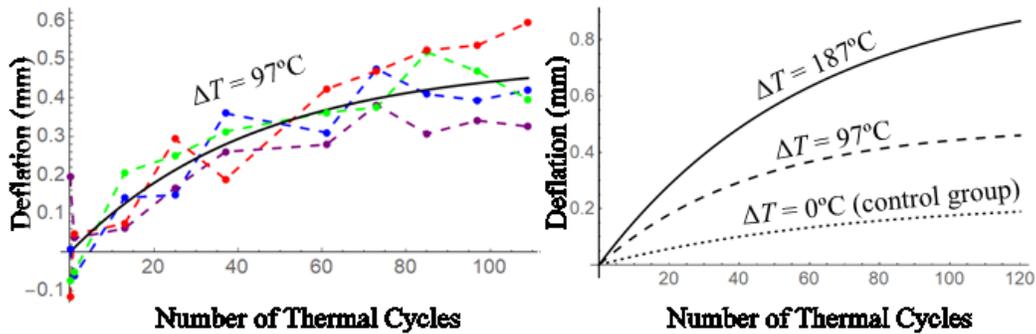

**Figure 2.** Results of Thermal Cycling. Left: For the $\Delta T$ = 97 °C case, four samples in dashed colors with the fitted function in solid black. Right: the fitted functions for three cases, $\Delta T$ =187 °C, 97 °C, and 0 °C (the control group that included all sample handling but no oven power).

The data were initially fitted with a double exponential following Chen et al. (2006),

$$x = a - c\exp(-bN) - (a-c)\exp(-dN)$$

where $x$ is amount of deflation (distance change to top of soil), $N$ is number of thermal cycles, and the rest are fitting parameters. It fits excellently for the $\Delta T$=187 °C case, suggesting that there are two compaction processes: a fast one for individual grain movements and a slow one for coordinated grain rearrangements, as Chen et al. suggested following Barker and Mehta (1993). However, a single exponential was adequate for all other cases by setting $a=c$ and $d=0$. Although this fit was slightly worse for the ΔT=187 °C case it removed degeneracy in the curve fitting for all the other cases, enabling more meaningful comparison of the fitting constants. This comparison of fitting constants is shown in Figure 3.

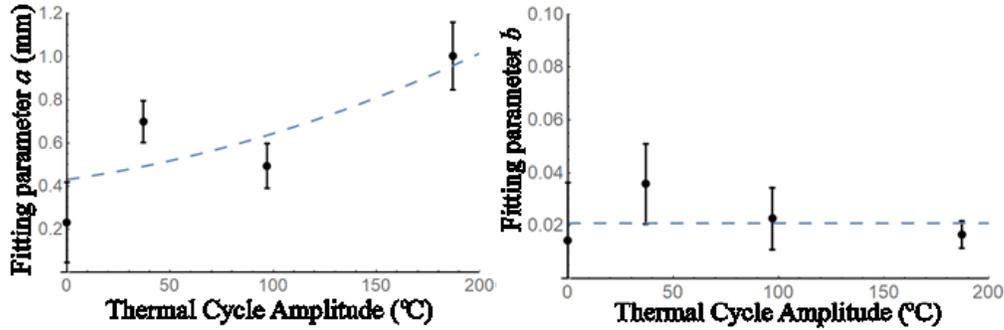

Figure 3. Exponential fitting parameters for soil deflation using the first method of data reduction. Error bars are two standard deviations.

The data from the ΔT=67 °C case could not be used in this first analysis because the data were too noisy. A second data reduction was performed, removing only sample curves that showed sudden step-function offsets (signs of vibrational shock) and further removing *individual* data points that were randomly out-of-family due (presumably) to misalignment of the displacement sensor, which was an easy error to make. This allowed the ΔT=67 °C case to be included, albeit with large error bars because only three sample cups were free of abrupt deflation events indicative of shocks. The results are shown in Figure 4. In both data reductions a quadratic equation was used to fit $a$ vs. $\Delta T$. For Figure 4 (left) the dashed line is

$$a = 0.289 + 2.13 \times 10^{-3}\, \Delta T + 6.42 \times 10^{-6} (\Delta T)^2$$

showing that soil becomes more compacted where the thermal wave amplitude is higher. The error bars are small enough that the trend lines are statistically significant. The $b$ parameter is consistent with a constant value.

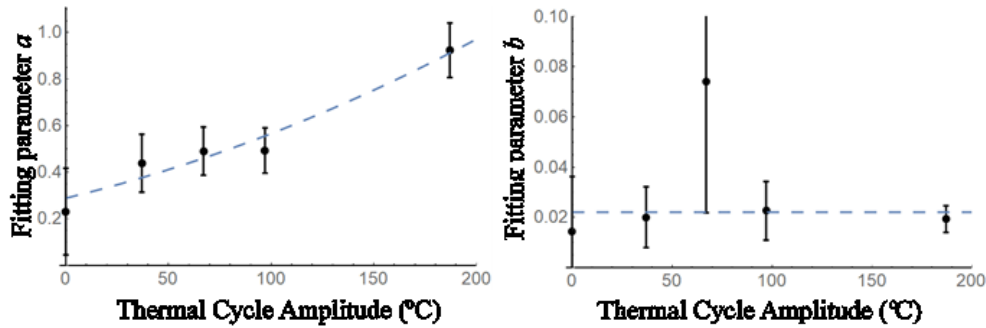

Figure 4. Exponential fitting parameters for soil deflation using the second method of data reduction. Error bars are two standard deviations.

## DISCUSSION:

While prior experiments (Clegg, 2009; Gamsky and Metzger, 2010) had already indicated that thermal cycling can compact lunar soil, these new experiments successfully eliminated (to 8.9 standard deviations) the effect of unwanted experimental vibration. This now proves effectively that it was in fact the thermal cycling, and not incidental experimental vibrations, that compacted the soil in these experiments. The new results also demonstrate a temperature trend that agrees with the Diviner Radiometer measurements as interpreted by thermal thermal modeling, that the soil is more compacted where the thermal wave has a higher amplitude. The experiments lend confidence that the interpretation of Diviner by Hayne (2015) is correct.

This small number of thermal cycles (~120) simulates less than 10 years of geological time. The deflation of the $\Delta T=0$ °C (control) set and the $\Delta T=187$ °C set over this short time correspond to average density changes of 0.26% and 1.3%, respectively, changing their mean bulk densities from the initial 1.6 g/cm$^3$ to the (calculated) final values of 1.604 and 1.621 g/cm$^3$, respectively. While this is small, the authors believe it underrepresents compaction in the lunar environment because of the Janssen Effect, in which bridges of grains form across the container, supporting vertical stress so the grains are better "locked" into place. This would retard compaction dynamics in the beakers used in these experiments, but not in the lunar environment where there are no container walls. There has been much progress in recent years in the theoretical statistical mechanics of idealized granular packings undergoing dynamics inside containers. Future work could develop a theory based on that theoretical progress to extrapolate these experiments to the lunar environment. Also, new measurements should be performed that seek to minimize the container effect by using large pans instead of beakers and/or by constructing broad "mesas" of soil bounded by slopes instead of containers.

One question about thermal cycling is whether the thermal wave penetrates deeply enough to explain the density at depth. The thermal wave is less than 2 °C at 50 cm depth (Carrier et al., 1991, Fig. 3.9). This seems small, but the average linear thermal expansion of basalt (a major component of lunar mare soils) is $5.4 \times 10^{-6}$ /°C (Dane, 1942). The mass-averaged particle size in lunar soil is about 60 µm in diameter, so at 50 cm depth it will expand via the temperature change by $5.4 \times 10^{-4}$ µm. Measurements of force/displacement curves on individual lunar soil and simulant particles by Cole and Peters (2008) shows the normal contact stiffness to be between 0.2 and 2 MN/m. Thus, the thermally-induced expansion (divided by two for radius rather than diameter) will produce between 65 and 650 µN of contact force. There will be about 290 million such particles on a square meter cross section of the lunar soil, so the total stress induced by thermal expansion will be between 18.8 and 188 kPa at that depth. The overburden in lunar gravity is only ~1.4 kPa at that depth, so thermal cycling can produce 13 to 130 times the overburden stress, easily overconsolidating the soil. At 10 cm depth the thermal stresses are 3,000 to 30,000 times the overburden. Perhaps more importantly, every distinct layer in the soil column was originally the top layer. If thermal cycling operates on a faster timescale than the blanketing of additional impact ejecta, then it can explain every layer's compaction. The exception would be the ejecta deposits on large crater rims where they are much thicker than the thermal wave. These may remain uncompacted below where the wave has tapered off. This seems in agreement with observations during the Apollo missions when penetrometers were easily inserted into the rims of large impact craters, but not elsewhere.

It is also possible that there is a synergistic relationship between thermal cycling and vibrating lunar soil. Perhaps thermal cycling weakens the fabric and prevents soil ageing (Schmertmann, 1991), which ordinarily strengthens soil over time. Then when vibrations arrive, the soil that could not age because of thermal cycling more easily collapses to a denser state. Also, it may be that the abrupt deflation events seen in these experiments are not the result of external shocks but are random, internal rearrangements of grains that set off small internal shocks that precipitate avalanches. These relationships need to be explored further. In any case, this synergy hypothesis does not detract from the interpretation that lunar soil will be less compacted in the polar regions than at mid latitudes.

## CONCLUSIONS

New experiments have shown the compaction of lunar soil does occur by thermal cycling (perhaps in conjunction with vibrations) and that less compaction occurs where the thermal wave amplitude is smaller. This suggests soil could be less compacted at high latitudes and remain very loose in PSRs. This is the only known compaction mechanism that can compact soil with latitude-dependence. These experiments support the indications of latitude dependence in Diviner data. A roving vehicle intended to operate in polar regions may need to be designed for driving in soil that is less compacted than was experienced in prior lunar surface missions.